\documentclass{aa}

\usepackage{graphicx}
\usepackage{txfonts}
\usepackage{hyperref}
\usepackage{color}
\def\Mjup{\hbox{$\thinspace M_{\mathrm{J}} \thinspace$}}
\def\Msun{\hbox{$\thinspace M_{\odot} \thinspace$}}
\def\Rsun{\hbox{$\thinspace R_{\odot} \thinspace$}}

\def\kms{\hbox{$\thinspace {\mathrm{km~s^{-1}}}$}}
\def\ms{\hbox{$\thinspace {\mathrm{m~s^{-1}}}$}}

\def\ALi{\hbox{$\thinspace A (\mathrm{Li})$}}

\def\au{\hbox{$\thinspace \mathrm{au}$}}
\def\sjit{\hbox{$\sigma_{\mathrm{jitter}}$}}
\def\srv{\hbox{$\sigma_{\mathrm{RV}}$}}
\def\sbs{\hbox{$\sigma_{\mathrm{BIS}}$}}

\def\starA{HD~103485\thinspace}
\def\starB{BD+03~2562\thinspace}
\def\shk{\hbox{$\thinspace S_{\mathrm{HK}}^{\mathrm{inst}}$}}
\newcommand\sprot{P_\mathrm{rot}}
\newcommand\spsini{P_\mathrm{rot} (\sin\,i_{\star})^{-1}}

\begin{document} 
 \title{Tracking Advanced Planetary Systems (TAPAS) with HARPS-N. V.  
 \thanks{Based on observations obtained with the Hobby-Eberly Telescope, which is a joint project of the University of Texas at Austin, the Pennsylvania State University, 
Stanford University, Ludwig-Maximilians-Universit\"at M\"unchen, and Georg-August-Universit\"at G\"ottingen.}
\thanks{Based on observations made with the Italian Telescopio Nazionale Galileo (TNG) operated on the island of La Palma by the Fundaci\'on Galileo Galilei of the INAF (Istituto Nazionale di Astrofisica) at the Spanish Observatorio del Roque de los Muchachos of the Instituto de Astrof\'{\i}sica de Canarias.}
}
   \subtitle{ {\bf A} Massive Jupiter orbiting the very low metallicity giant star BD+03 2562 {\bf and a possible planet around}  HD~103485.}
   \titlerunning{TAPAS V. .}
   \authorrunning{E. Villaver et al.}
 \author{E. Villaver \inst{1} \and
         A. Niedzielski \inst{2} \and
         A. Wolszczan \inst{3,4} \and               
         G. Nowak \inst{5,6,2} \and                                
         K. Kowalik \inst{7} \and    
         M. Adam\'ow \inst{8,2} \and    
         G. Maciejewski \inst{2} \and                                                  
         B. Deka-Szymankiewicz  \inst{2} \and
         J. Maldonado\inst{9}}
   \institute{Departamento de F\'{\i}sica Te\'orica, Universidad Aut\'onoma de
      Madrid, Cantoblanco 28049 Madrid, Spain.
   \email{Eva.Villaver@uam.es}\and
     Toru\'n Centre for Astronomy, Faculty of Physics, Astronomy and
   Applied Informatics, Nicolaus Copernicus University in Toru\'n,
   Grudziadzka~5, 87-100 Toru\'n, Poland.
   \email{Andrzej.Niedzielski@umk.pl}  \and
   Department of Astronomy and Astrophysics, Pennsylvania State
   University, 525 Davey Laboratory, University Park, PA 16802, USA \and
   Center for Exoplanets and Habitable Worlds, Pennsylvania State
   University, 525 Davey Laboratory, University Park, PA 16802, USA. \and
   Instituto de Astrof\'isica de Canarias, E-38205 La Laguna,
   Tenerife, Spain. \and
   Departamento de Astrof\'isica, Universidad de La Laguna, E-38206
   La Laguna, Tenerife, Spain. \and
   National Center for Supercomputing Applications, University of
   Illinois, Urbana-Champaign, 1205 W Clark St, MC-257, Urbana, IL 61801,
   USA \and
   McDonald Observatory and Department of Astronomy, University of
      Texas at Austin, 2515 Speedway, Stop C1402, Austin, Texas, 78712-1206,
   USA. \and
   INAF - Osservatorio Astronomico di Palermo, Piazza del Parlamento 1, I-90134 Palermo, Italy}

   \date{Received;accepted}

  \abstract
   {We present  two evolved stars (\starA and \starB) from the TAPAS (Tracking Advanced PlAnetary Systems) with HARPS-N project devoted to RV precision measurements  of identified candidates within the PennState - Toru\'n Centre for Astronomy Planet Search.} 
   {Evolved stars with planets are crucial to understand the dependency of the planet formation mechanism on the mass and metallicity of the parent star and to study star-planet interactions.}
   {The paper is based on precise radial velocity (RV) measurements. For \starA we collected 57 epochs
   over 3317 days with the Hobby-Eberly Telescope  and its High Resolution Spectrograph  and 18 ultra-precise HARPS-N data over 919 days. For \starB we collected 46 epochs of HET data over 3380 days and 19 epochs of HARPS-N data over 919 days.}
   {We present the analysis of the data and the search for correlations between the RV signal and stellar activity, stellar rotation and photometric variability. Based on the available data, we interpret the RV variations measured in both stars as Keplerian motion. Both stars have masses close to Solar (1.11\Msun \starA\thinspace and 1.14\Msun \starB), very low metallicities ([Fe/H$]= -0.50$ and $-0.71$ for \starA and \starB), and, both have Jupiter planetary mass companions ($m_2\sin i\thinspace=\thinspace7$  and 6.4 \Mjup for \starA and \starB resp.), in close to terrestrial orbits (1.4~au \starA and 1.3~au \starB), with moderate eccentricities ($e=0.34$ and 0.2 for \starA  and \starB). However, we cannot totally exclude that the signal in the case of HD~103485 is due to rotational modulation of active regions.}
   {Based on the current data, we conclude that BD+03~2562 has a bona fide planetary companion while for HD~103485 we cannot totally exclude that the best explanation for the RV signal modulations is not the existence of a planet but stellar activity.  If, the interpretation remains that both stars have planetary companions they represent systems orbiting very evolved stars with very low metallicities, a challenge to the conditions required for the formation of massive giant gas planets.}


   \maketitle
%

\section{Introduction}

With the discovery of 51~Peg \citep{Mayor1995} we arrived to the realisation that planet formation as we understood it had to be revised to account for the existence of "Hot Jupiters". Since then, every planetary system discovered has added to our understanding of the physics of planet formation (see e.g. \citealt{2007ARA&A..45..397U}). In this regard, planets orbiting evolved stars hold the key to several processes  related not only to how planet formation operates around stars more massive than the Sun but also to understand star-planet interactions \citep{Villaver2007, Villaver2014, Privitera2016c,Privitera2016a}. In this context, planets around evolved stars have revealed a lack of hot Jupiters that most likely reflect effects induced by stellar evolution \citep{VillaverLivio2009, Mustill2012,Villaver2014, Privitera2016b}. 

In the main sequence, the presence of giant planets has shown to be very sensitive to the metallicity 
[Fe/H] of the host star  \citep{Gonzalez1997, Santos2004,FischerValenti2005}. The precise functional form of the correlation still remain elusive, despite the fact  it is one of the fundamental parameters to help constrain the planet formation models
(see e.g. \citealt{Mordasini2012}). But most important, the evolved hosts of planets 
have shown to present some chemical peculiarities with respect to their main sequence counterparts (see e.g. \citealt{Pasquini1997, daSilva2006, Ghezzi2010,Maldonado2013,Mortier2013,Jofre2015,Jones2014,Reffert2015,Maldonado2016}).  In particular, the planet occurrence rate does seem to depend on both stellar mass and stellar metallicity  \citep{Maldonado2013} and cannot be explained by sample contamination \citep{Maldonado2016} as it has been argued by \cite{Reffert2015}.

The established picture of giant planet formation, the basis of the core accretion model \citep{Perri1974, Cameron1978,Mizuno1980,Pollack1996}, begins with the building of km-sized or larger planetesimals from the growth of 1$-$10~mm pebbles (see e.g. \citealt{Youdin2011,Simon2016}). The  planetesimals growth continues until a solid core, big enough for gravity to accrete gas from the protoplanetary disk, is formed. The metallicity dependency in the core accretion model thus comes from the need of a fast core growth before disk dissipation occurs \citep{Ida2004}. Furthermore, it has been shown that the alternative scenario, giant planet formation via gravitational instability in the protoplanetary disk \citep{Boss1997,Mayer2002, Boss2004} does not carry a metallicity dependency that can explain the observed relation.

Over the last $\approx 10$ years we have embarked in a quest for substellar/planetary companions to giant stars that started with the {PennState - Toru\'n Centre for Astronomy Planet Search} (PTPS,
\citealt{Niedzielski2007,  NiedzielskiWolszczan2008, Niedzielski2015b,
Niedzielski2016a}) program and has continued with the high precision RV follow up of previous selected PTPS candidates program {Tracking Advanced Planetary Systems (TAPAS) program with HARPS-N}
\citep{Niedzielski2015a,Adamow2015,Niedzielski2016b, Niedzielski2016c}. In this paper, we present the latest finding of our TAPAS program: two very evolved giant stars with very low metallicities that host {\bf a} massive "warm" Jupiters and a  possible one and thus represent rather extreme outliers to the general planet-metallicity relation.

The paper is organised as follows:  a summary of the observations,  radial velocity and activity measurements is given in Section \ref{observations} together with a description of the general procedure and the basic properties of the two stars; in Section \ref{rot} and \ref{phot} we show the analysis of the stellar rotation and photometry together with a discussion of the activity indicators; in Section \ref{results-g} we present the Keplerian analysis of the radial velocity measurements, and our results are summarised and
further discussed in Section \ref{conclusions}.


\section{Observations, radial velocities, line bisectors and activity indicators\label{observations}}

HD~103485 (BD+02~2493) and BD+03~2562 (TYC~0276-00507-1)  belong to a sample of  about 300 planetary or brown dwarf (BD) candidates
identified from a sample of about 1000 stars searched for radial velocity (RV) variations with the
9.2m Hobby-Eberly Telescope (HET, \citealt{Ramsey1998}).  The full sample have been monitored  since 2004 using the High-Resolution Spectrograph  (HRS, 
\citealt{Tull1998})  at HET within the PTPS
program. Targets were selected for a more intense precise RV follow-up within 
the TAPAS program with the High
Accuracy Radial velocity Planet Searcher in the North hemisphere (HARPS-N,
\citealt{Cosentino2012}).

The spectroscopic observations presented in this paper are thus a combination of data taken with
the  HRS at HET in the queue scheduled mode \citep{Shetrone2007},
and with HARPS-N at the 3.58 meter Telescopio Nazionale Galileo (TNG). 

For  HET HRS spectra we  use a combined gas-cell  \citep{MarcyButler1992,
Butler1996}, and  cross-correlation \citep{Queloz1995, Pepe2002} method for
precise RV and spectral bisector inverse slope (BIS) measurements,
respectively.  The implementation of this technique to our data is described in
\cite{Nowak2012} and \cite{ Nowak2013}. 

HARPS-N radial velocity measurements and their uncertainties as well as BIS measurements were
obtained with the standard user pipeline, which is based on the weighted CCF
method \citep{1955AcOpt...2....9F, 1967ApJ...148..465G, 1979VA.....23..279B,
Queloz1995, Baranne1996, Pepe2002}, using the simultaneous Th-Ar calibration
mode of the spectrograph  and the K5 cross-correlation mask. 

A summary of the available data for HD~103485 and BD+03~2562 
is given in Tables \ref{Parameters1} and \ref{Parameters2} respectively.

\begin{table}
\centering
\caption{Summary of the available data on HD 103485}
\begin{tabular}{lll}
\hline
Parameter & value & reference\\
\hline
\hline
$V$  [mag]&  8.28$\pm$0.01 &  \cite{Hog2000} \\
$B-V$ [mag] &  1.56$\pm$ 0.03 & \cite{Hog2000} \\ 
$(B-V)_0$ [mag] & 1.395& \cite{Zielinski2012}\\
$M_\mathrm{V}$ [mag] & -2.51& \cite{Zielinski2012} \\
 $T_{\mathrm{eff}}$ [K] & 4097$\pm$20 & \cite{Zielinski2012} \\
 $\log g$ & 1.93$\pm$0.08& \cite{Zielinski2012} \\
$[Fe/H]$ &    -0.50$\pm$0.09 & \cite{Zielinski2012}\\
RV $[\kms]$ &  27.56$\pm$0.08 & \cite{Zielinski2012} \\
$v_{\mathrm{rot}} \sin i_{\star}$ $[\kms]$ & 2.9$\pm$0.4 & \cite{Adamow2014} \\
$\ALi $& $ -0.84$ &  \cite{Adamow2014} \\
$[$O/H$]$ & -0.40$\pm$0.31& \cite{Adamow2014}\\
$[$Mg/H$]$ & -0.27$\pm$0.15 & \cite{Adamow2014}\\
$[$Al/H$]$ & -0.10$\pm$0.09  & \cite{Adamow2014}\\
$[$Ca/H$]$ & -0.54$\pm$0.18 & \cite{Adamow2014}\\
$[$Ti/H$]$ & -0.07$\pm$0.25& \cite{Adamow2014}\\
\hline
$M/M_{\odot}$ & 1.11$\pm$0.21 & \cite{Adamczyk2015}\\
$\log L/L_{\odot}$ & 2.51$\pm$0.13 & \cite{Adamczyk2015}\\
$R/R_{\odot}$ & 27.37$\pm$6.69 & \cite{Adamczyk2015}\\
$\log \mathrm{age}$ [yr]& 9.79$\pm$0.25& \cite{Adamczyk2015} \\
$d$ [pc] & 1134$\pm$ 145 & calculated from M$_{V}$\\
$V_{\mathrm{osc}}$ [$\ms$] & 68.2$^{+52.1}_{-28.2}$ & this work\\
$P_{\mathrm{osc}}$ [d] & 2.2$^{+2.0}_{-1.1}$ & this work\\
$P_{\mathrm{rot}}/ \sin i_{\star}$ [d] & 477$\pm$134 & this work\\ 
\hline
\hline
\end{tabular}
\label{Parameters1}
\end{table}


\begin{table}
\centering
\caption{Summary of the available data on BD+03 2562}
\begin{tabular}{lll}
\hline
Parameter & value & reference\\
\hline
\hline
$V$  [mag]& 9.58$\pm$0.01 &  \cite{Hog2000} \\
$B-V$ [mag] & 1.27 $\pm$ 0.09 & \cite{Hog2000} \\ 
$(B-V)_0$ [mag] & 1.38& \cite{Zielinski2012}\\
$M_\mathrm{V}$ [mag] & -2.51& \cite{Zielinski2012} \\
$T_{\mathrm{eff}}$ [K] & 4095$\pm$20 & \cite{Zielinski2012} \\
$\log g$ & 1.89$\pm$0.10& \cite{Zielinski2012} \\
$[Fe/H]$ &    -0.71$\pm$0.09 & \cite{Zielinski2012}\\
RV $[\kms]$ &  50.88$\pm$0.06 & \cite{Zielinski2012} \\
$v_{\mathrm{rot}} \sin i_{\star}$ $[\kms]$ & 2.7$\pm$0.3 & \cite{Adamow2014} \\
$\ALi $& $-0.56$ &  \cite{Adamow2014} \\
$[$O/H$]$ & -0.23$\pm$0.22& \cite{Adamow2014}\\
$[$Mg/H$]$ & -0.01$\pm$0.13 & \cite{Adamow2014}\\
$[$Al/H$]$ & -0.21$\pm$0.10  & \cite{Adamow2014}\\
$[$Ca/H$]$ & -0.68$\pm$0.18 & \cite{Adamow2014}\\
$[$Ti/H$]$ & -0.34$\pm$0.24& \cite{Adamow2014}\\
\hline
$M/M_{\odot}$ & 1.14$\pm$0.25 & \cite{Adamczyk2015}\\
$\log L/L_{\odot}$ & 2.70$\pm$0.14 & \cite{Adamczyk2015}\\
$R/R_{\odot}$ & 32.35$\pm$8.82 & \cite{Adamczyk2015}\\
$\log \mathrm{age}$ [yr]& 9.72$\pm$0.28& \cite{Adamczyk2015}\\
$d$ [pc] &  2618 $\pm$ 564 & calculated from M$_{V}$\\
$V_{\mathrm{osc}}$ [$\ms$] & 102.9$^{+89.9}_{-45.4}$ & this work\\
$P_{\mathrm{osc}}$ [d] & 2.9$^{+2.0}_{-1.1}$ & this work\\
$P_{\mathrm{rot}} / \sin i_{\star}$ [d] & 606$\pm$179 & this work\\ 
\hline
\hline
\end{tabular}
\label{Parameters2}
\end{table}

\subsection{RV and BIS}
The 57 epochs of HET/HRS data for HD 103485 
show RV variations of $553 \ms$ with average uncertainty of $5.6 \ms$ and BIS variations of $104\ms$ with an average uncertainty of $16 \ms$. No correlation between RV and BIS exists (Pearson's r=0.09). HARPS-N RV 18 epochs of data show an amplitude of $553 \ms$ (average uncertainty of $1.7 \ms$). The BIS shows a peak-to-peak amplitude of $85 \ms$ and no correlation with RV (r=-0.02).
Lomb-Scargle (LS) periodogram analysis \citep{1976ApJSS..39..447L, 1982ApJ...263..835S, 1992nrfa.book.....P} of combined HET/HRS and HARPS-N RV data reveals a strong periodic signal in RV at 557~days.

For BD+03 2562 
the 46 epochs of HET/HRS data  show RV variations of $575\ms$ with average uncertainty of  $7\ms$ and BIS variations of $136\ms$ with average uncertainty of $22\ms$. The 
19 epochs of HARPS-N data show RV amplitude of $444\ms$ and average uncertainty of $2.1\ms$. The BIS shows a peak-to-peak amplitude of $58\ms$. There is no correlation between RV and BIS in either HET/HRS data (r=0.23) or HARPS-N (r=0.13) data. A strong periodic signal in RV at 482 days appears in the LS periodogram analysis of combined HET/HRS and HARPS-N RV data. 

HET/HRS and HARPS-N BIS have to be considered separately due to their different definition (see \citealt{Niedzielski2016b} for more details). The RV and BIS data for both stars are presented in Tables \ref{HETdata1}, \ref{HETdata2}, \ref{HARPSdata1} and \ref{HARPSdata2}.

\subsection{Activity indicators: the Ca H$\&$K lines}

The Ca II H, and K line profiles (see \citealt{ Noyes1984, Duncan1991}) 
and  the reversal profile, typical for active stars
\citep{EberhardSchwarzschild1913}
are widely accepted as stellar activity indicators.
The Ca~II H
and K lines are only available to us in the TNG HARPS-N spectra. The signal-to-noise of our red giants in that spectral 
range is low, 3-5 in this particular case, but we found no trace of  reversal. To quantify the observations, we
calculated an instrumental $\shk$ index according to the prescription of \cite{Duncan1991} 
for  HARPS-N data. The $\shk$ index for TNG HARPS-N spectra was calibrated 
to the Mt Wilson scale with the formula given by \cite{2011arXiv1107.5325L}.
For HD 103485, 
we obtained a value of  $0.20\pm0.05$ and
for BD+03 2562 
 of  $0.14\pm0.07$, rather typical values  for non-active  stars. 

The $\shk$ for both stars show no 
statistically significant correlation with the RV ($r=-0.41$  and $r=-0.25$, respectively). 
In order to dig dipper into the posible correlation we have performed further statistical Bayesian tests following the prescription given in \cite{Figueira2016}.
For \starA the Pearson's coefficient of the data is $0.408$ with a $0.093$ 2-sided p-value and the Spearman's rank coefficient is 0.514 (0.029 2-sided p-value). The distribution of the parameter of interest, $\rho$, characterizing the strength of the correlation is $0.327$ with a standard deviation of $0.187$  and 95\% credible interval [-0.049  0.663].  For \starB the Pearson's coefficient  is $0.254$, $0.310$ 2-sided p-value and the Spearman's rank coefficient  $0.051$ with a $0.841$ 2-sided p-value.  $\rho$ = 0.2 with a $0.197$  and 95\% credible interval  [-0.201  0.559]. For both stars a correlation is not conclusively seen and seems unlikely, with the 95\% credible interval lower limit being above $\rho$= 0. However, a note of warning is in place here given that in the case of rotational modulation, it is expected to have non-linear relations between RV and activity. This is caused by a phase shift between the activity maximum (that occurs when the active regions are at the centre of the disk) and the maximum RV effect, that happens at a phase of  $\approx$ 60$^{\circ}$.
 
Thus, we conclude that the Ca~II H and K line profile analysis reveals that, over the period
covered by TNG observations, both giants are quite inactive and there is no trace 
of activity influence upon the observed RV variations.

\subsection{Activity indicators: $\mathbf{H\alpha}$ analysis}

\cite{2007A&A...469..309C} showed that the calcium and hydrogen lines indices do not always correlate and cannot be used interchangeably as activity indicators. We thus measured the H$\alpha$ activity index ($I_{H\alpha}$) in both HET/HRS and TNG/HARPS-N spectra, following the procedure described in detail by \cite{2013AJ....146..147M}, which based on the approach presented by \cite{2012A&A...541A...9G} and \citet[][and references therein]{2013ApJ...764....3R}. 
We also measured the index in the Fe~I 6593.883~{\AA} control line ($I_{Fe}$) which is insensitive to stellar activity to take possible instrumental effects into account. Moreover, in the case of HET/HRS spectra, that may still contain weak $\mathrm{I_{2}}$ lines in the wavelength regime relevant to the H$\alpha$ and Fe~I 6593.883~{\AA} lines, we also measured the H$\alpha$ and Fe~I indices for the iodine flat-field spectra ($I_{\mathrm{I_{2}}, H\alpha}$ and $I_{\mathrm{I_{2}}, {Fe}}$ respectively).

\subsubsection{BD+03 2562}
The marginal rms variations of the $I_{\mathrm{I_{2}}, H\alpha}$ = 0.11\% and $I_{\mathrm{I_{2}}, {Fe}}$ = 0.33\% in comparison to the relative scatter of $I_{H\alpha,HRS}$ = 3.03\% and $I_{Fe,HRS}$ = 1.08\% assure us of the negligible contribution of the weak iodine lines to H$\alpha$ and Fe~I 6593.883~{\AA} line indices measured from HET/HRS spectra of BD+03 2562. The rms variation of $H\alpha$ activity index measured from 18 TNG/HARPS-N spectra is slightly larger than the one measured from the HET/HRS spectra ($I_{H\alpha,HARPS-N}$ = 3.68\%). The rms variation of TNG/HARPS-N Fe~I 6593.883~{\AA} line index is more than two times larger than that measured from HET/HRS spectra ($I_{Fe,HARPS-N}$ = 2.39\%). The larger rms variations of the line indices measured from the TNG/HARPS-N spectra compared to the HET/HRS spectra measurements might be a consequence of the lower SNR of the TNG/HARPS-N spectra (40--70), compared to that of the HET/HRS spectra (120--220). There is no correlation between neither HET/HRS $I_{H\alpha,HRS}$ and RVs (the Pearson coefficient, $r$ = 0.27), nor between TNG/HARPS-N $I_{H\alpha,HARPS-N}$ and RVs, (the same value of $r$ = 0.27). There is no any significant signal in the Lomb-Scargle periodograms of BD+03 2562 $H\alpha$ indices.

\subsubsection{HD~103485}

\begin{figure}[t]
\centerline{\includegraphics[angle=0,width=\columnwidth]{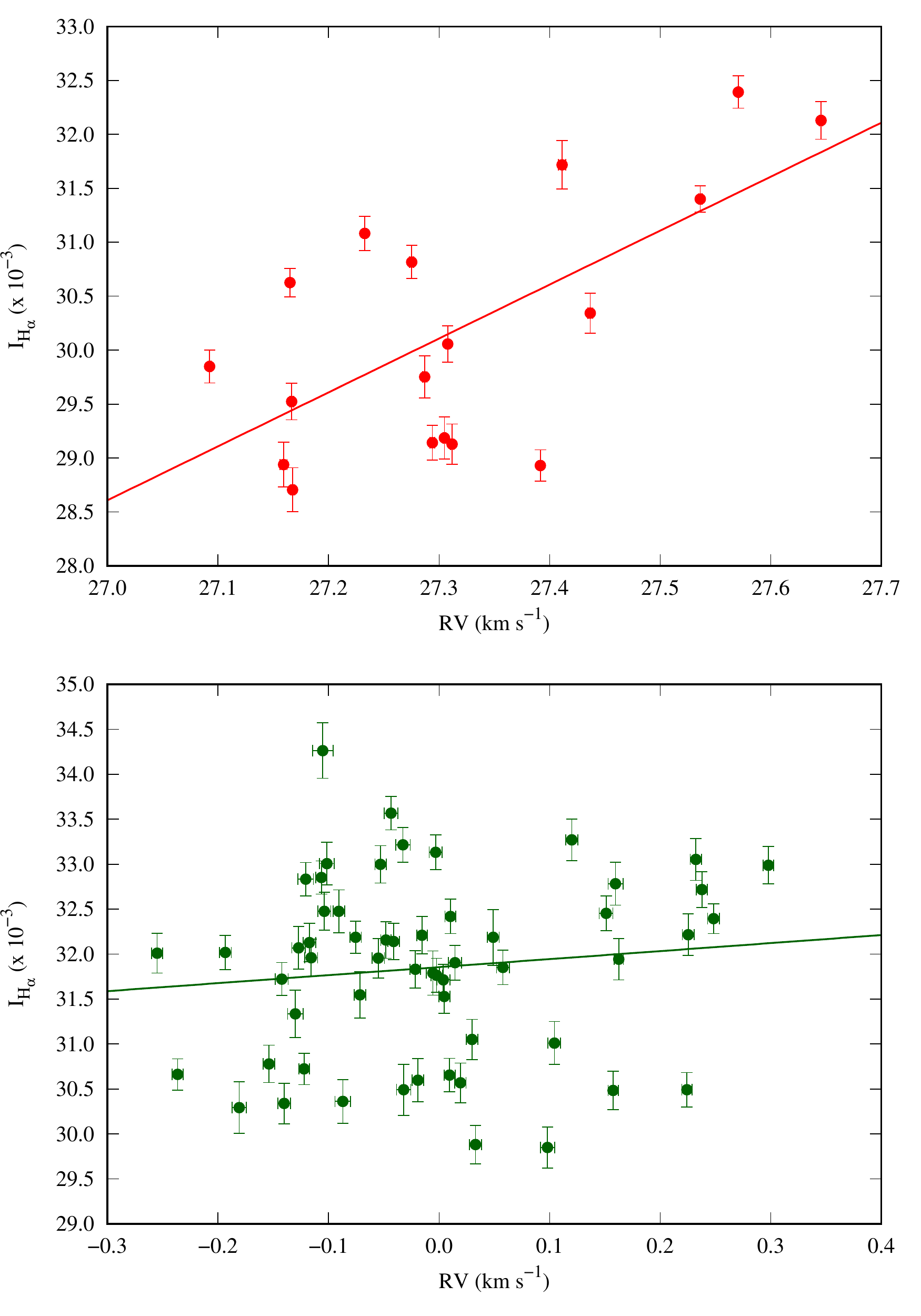}}
\caption{H$\alpha$ activity index of HD~103485 measured from TNG/HARPS-N spectra (top panel) and from HET/HRS ones (lower panel). Solid lines presents linear fits to the data. Although H$\alpha$ activity index measured from HET/HRS does not present correlation with HET/HRS radial velocities, the one measured form TNG/HARPS-N spectra shows clear correlation with TNG/HARPS-N radial velocities.
\label{hd103485-rv_vs_hai}}
\end{figure}

\begin{figure}[t]
\centerline{\includegraphics[angle=0,width=\columnwidth]{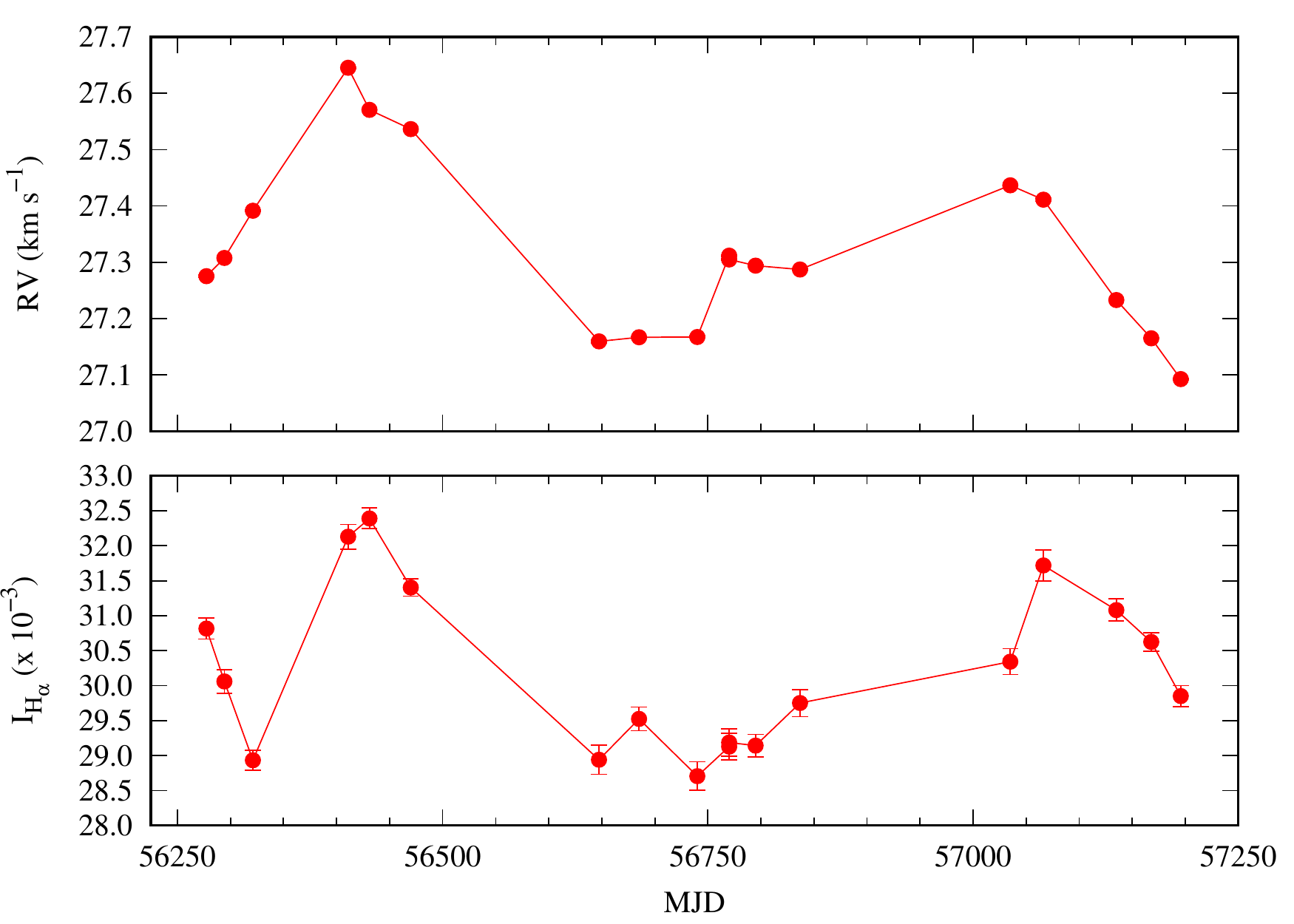}}
\caption{The TNG/HARPS-N radial velocity and H$\alpha$ activity index curves of HD~103485.
\label{hd103485-rv-hai-tng_harpsn}}
\end{figure}

\begin{figure}[t]
\centerline{\includegraphics[angle=0,width=\columnwidth]{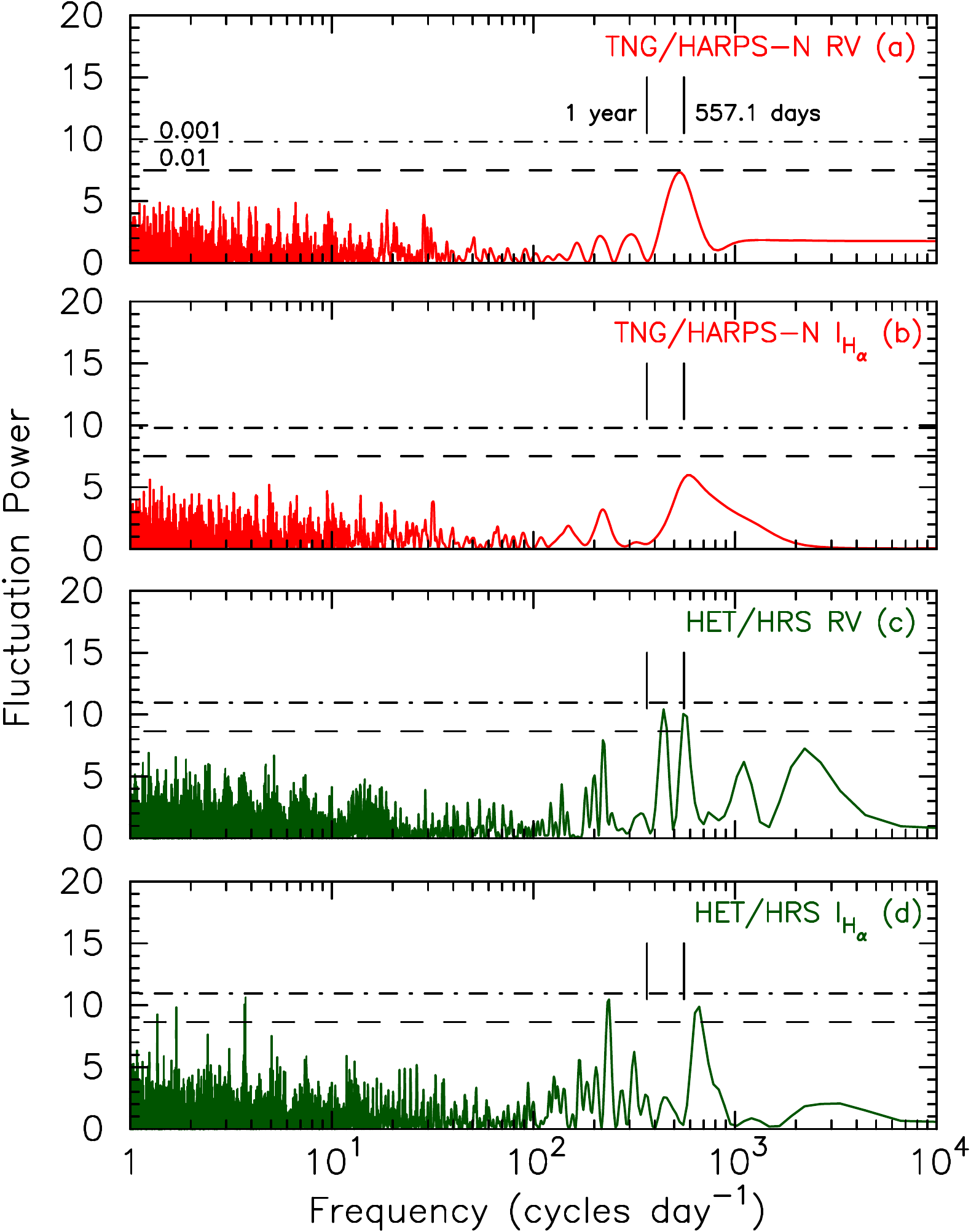}}
\caption{LS periodograms of (a) TNG/HARPS-N RVs, (b) TNG/HARPS-N H$\alpha$ activity index, (c) HET/HRS RVs, and (d) HET/HRS H$\alpha$ activity index of HD~103485. The levels of FAP = 1.0\% and 0.1\% are shown.
\label{hd103485-lsp}}
\end{figure}

The iodine flat-field HET/HRS spectra of HD~103485 show marginal rms variations ($I_{\mathrm{I_{2}}, H\alpha}$ = 0.1\% and $I_{\mathrm{I_{2}}, {Fe}}$ = 0.13\%) in the H$\alpha$ and Fe~I 6593.883~{\AA} indices. Comparing to the relative scatter of H$\alpha$ and Fe~I indices ($I_{H\alpha,HRS}$ = 3.16\% and $I_{Fe,HRS}$ = 0.79\%) we are assured of the negligible contribution of the weak iodine lines to H$\alpha$ and Fe~I 6593.883~{\AA} line indices. The Pearson coefficient between HET/HRS H$\alpha$ index and RVs is $r$ = 0.11. The rms variations of H$\alpha$ and Fe~I 6593.883~{\AA} line indices measured from 18 TNG/HARPS-N spectra are similar to those measured for BD+03 2562: $I_{H\alpha,HARPS-N}$ = 3.86\% and $I_{Fe,HARPS-N}$ = 2.73\%. However, the Pearson coefficient between TNG/HARPS-N H$\alpha$ index and RVs, $r$ = 0.66, while the critical value of the Pearson correlation coefficient at the confidence level of 0.01, $r_{16,0.01}$ = 0.59. On the other hand, this value ($r$ = 0.66) is lower than the critical value of the Pearson correlation coefficient at the confidence level of 0.001 ($r_{16,0.001}$ = 0.71). Given the small number of epochs the correlation may very well be spurious.
The H$\alpha$ indices versus HET/HRS and TNG/HARPS-N radial velocities are presented on Figure~\ref{hd103485-rv_vs_hai}. Figure~\ref{hd103485-rv-hai-tng_harpsn} presents the TNG/HARPS-N radial velocity and H$\alpha$ activity index curves of HD~103485. Figure~\ref{hd103485-lsp} presents LS periodograms of TNG/HARPS-N and HET/HRS radial velocities and H$\alpha$ indices.

\subsection{Wavelength dependence of the radial velocity signal}
 As both HD~103485 and BD+03 2562 exhibit significant scatter both in radial velocities and H$\alpha$ activity indices, the unambiguous interpretation of their origin is very difficult. Therefore, we analysed the wavelength dependence of the radial velocity peak-to-peak amplitude ($A$). The value of $A$ should be constant in the case of spectral shifts induced by the gravitational pull of the companion. In the case of a RV signal generated by the rotation of a spotted stellar photosphere, the value of $A$ should decrease with increasing wavelength as the temperature difference between the stellar photosphere and a stellar spot decreases at longer wavelengths \citep[see e.g. ][]{Saar1997,Hatzes2002, 2007A&A...473..983D}.

\begin{figure}[t]
\centerline{\includegraphics[angle=0,width=\columnwidth]{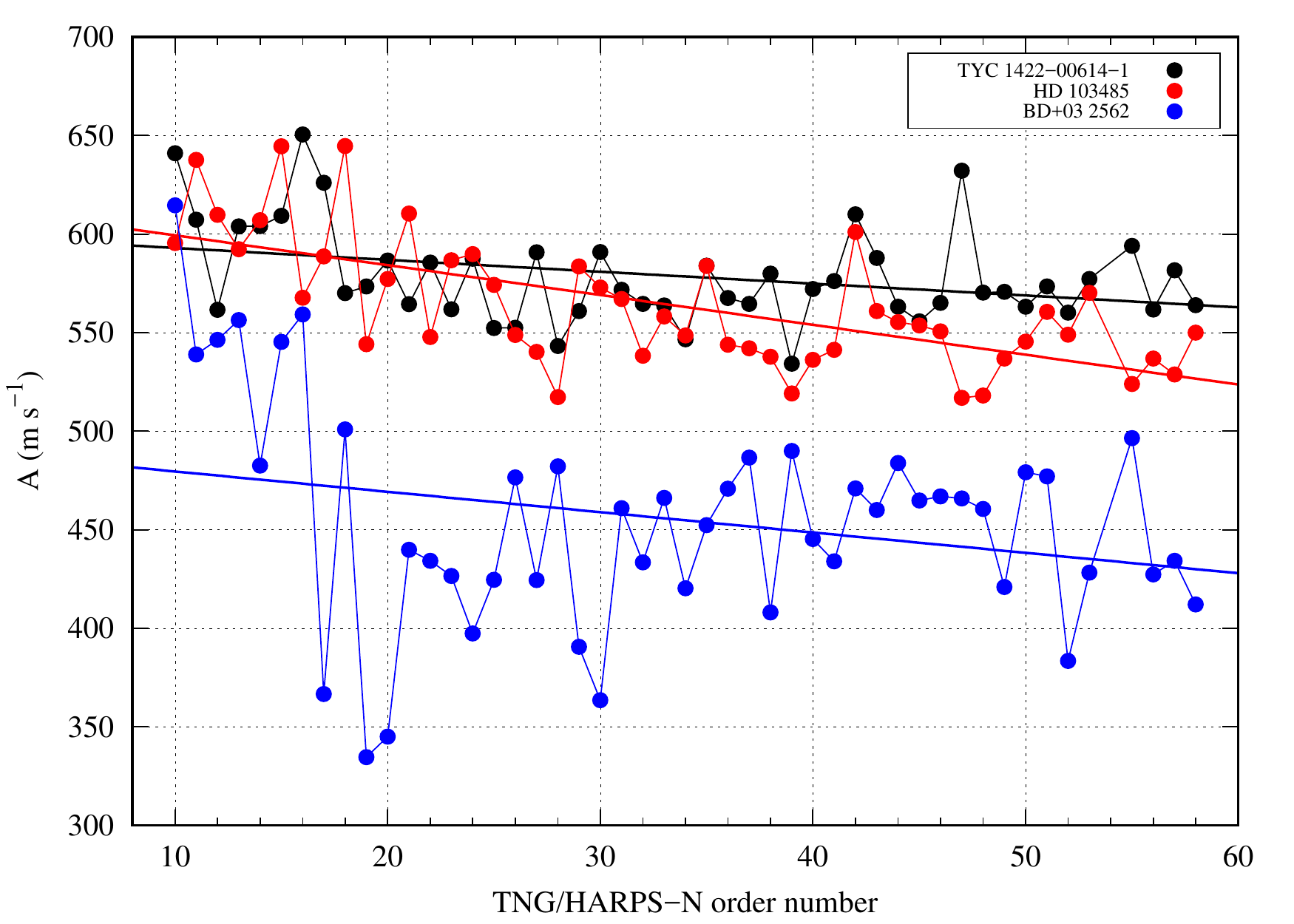}}
\caption{Radial velocity peak-to-peak amplitude ($A$) as a function of TNG/HARPS-N order number.
\label{rrv_vs_rno}}
\end{figure}

Figure~\ref{rrv_vs_rno} shows, $A$, the peak-to-peak RV amplitude as a function of the TNG/HARS-N order number for HD~103485, BD+03 2562, and the multiple planetary host PTPS target TYC 1422-00614-1 presented in \cite{2015A&A...573A..36N} (TAPAS-I paper). TYC 1422-00614-1 does not show any significant stellar activity related to any of the two signals reported in its radial velocity curve. Therefore, it is a good benchmark to test the wavelength dependence of the radial velocity signal peak-to-peak amplitudes of the stars in this paper HD~103485 and BD+03 2562. As shown in Figure~\ref{rrv_vs_rno}, both HD~103485 and BD+03 2562 show chromatic dependence of $A$, although its unambiguous interpretation is not straightforward, especially if we note that the peak-to-peak RV amplitude is systematically higher in TNG/HARPS-N orders 10--16.

Equation (5) of \cite{2007A&A...473..983D} gives the relation between the peak-to-peak amplitude of the RV variation ($A$), the projected rotation velocity of the star ($v_{\mathrm{rot}} \sin i_{\star}$) and the fraction of the visible hemisphere of the star that might be covered by the spot (parameter $f_{r}$, see \citealt{2007A&A...473..983D} for its definition and relation to the fraction of the projected area covered by the spot, $f_{p}$, on the 2D stellar disk used by other authors). Using the above mentioned equation we computed the parameter $f_{r}$ for HD~103485 and BD+03 2562. As inputs in equation (5) we used the values of the projected rotation velocities of both stars from Tables~1 and 2 and the values of the $K$ semi-amplitudes from Tables 3 and 4 ($A$ is  $\approx$ $2K$). For HD~103485 we obtained  $f_{r} = 5.93$ \% and for BD+03 2562  $f_{r} = 5.71$\%. Then, using equation (6) of \cite{2007A&A...473..983D}, that gives the relation between the peak-to-peak amplitude of the bisector inverse slope peak-to-peak variation ($S$), the parameter $f_{r}$, $v_{\mathrm{rot}} \sin i_{\star}$ and the instrumental width of the spectrograph ($v_0 = 3$ \kms for both HARPS and HAPRS-N) we computed the values of $S$ for both of our targets. We obtain $S = 96$ \ms~for HD~103485 and $S = 75.5$ \ms for BD+03 2562. The computed value of the peak-to-peak amplitude of bisector inverse slope for HD 103458 is consistent with its peak-to-peak amplitude of TNG/HARPS-N BIS ($85$ \ms, see section 2.1.), while in the case of BD+03~2562, it is significantly higher (see section 2.1. TNG/HARPS-N BIS for  BD+03~2562 is 58 \ms). We have to remember though, that equations (5) and (6) were derived for main sequence K5 type stars and HD~103458 and BD+03~2562 are giant stars.

\section{Stellar rotation and solar-like oscillations\label{rot}}
The one sigma limit of the rotation period ($\spsini$) is equal to 477 $\pm$ 134 days for HD~103485 and 606 $\pm$ 179 days for BD+03~2562. Thus, the true rotation period ($\sprot$) of HD~103485 at one sigma is then lower than 611 days and $\sprot$ is lower than 785 days for BD+03~2562. Based on the upper limits of the rotation periods of HD~103485 and BD+03~2562 we then cannot exclude that the signals in their radial velocity curves are generated by rotational modulation of active photospheres.

The amplitudes ($V_{osc}$) and periods ($P_{osc}$) of solar-like oscillations computed using equations (7) and (10) of \cite{1995A&A...293...87K} are equal to $68.2^{+52.1}_{-28.2}$ \ms and $2.2^{+2.0}_{-1.1}$ days for HD~103485 and to $102.9^{+89.9}_{-45.4}$ \ms and $2.9^{+3.2}_{-1.7}$ days for BD+03~2562. Computed values of solar-like oscillations amplitudes are consistent with the values of stellar jitter ($\sigma_{jitter}$) and post-fit rms ($\mathrm{RMS}$) presented in Tables~3 and~4. Both stars exhibit extremely high stellar jitter.

\section{Photometry and discussion of activity indicators\label{phot}}

\begin{figure}
   \centering
   \includegraphics[width=0.5\textwidth]{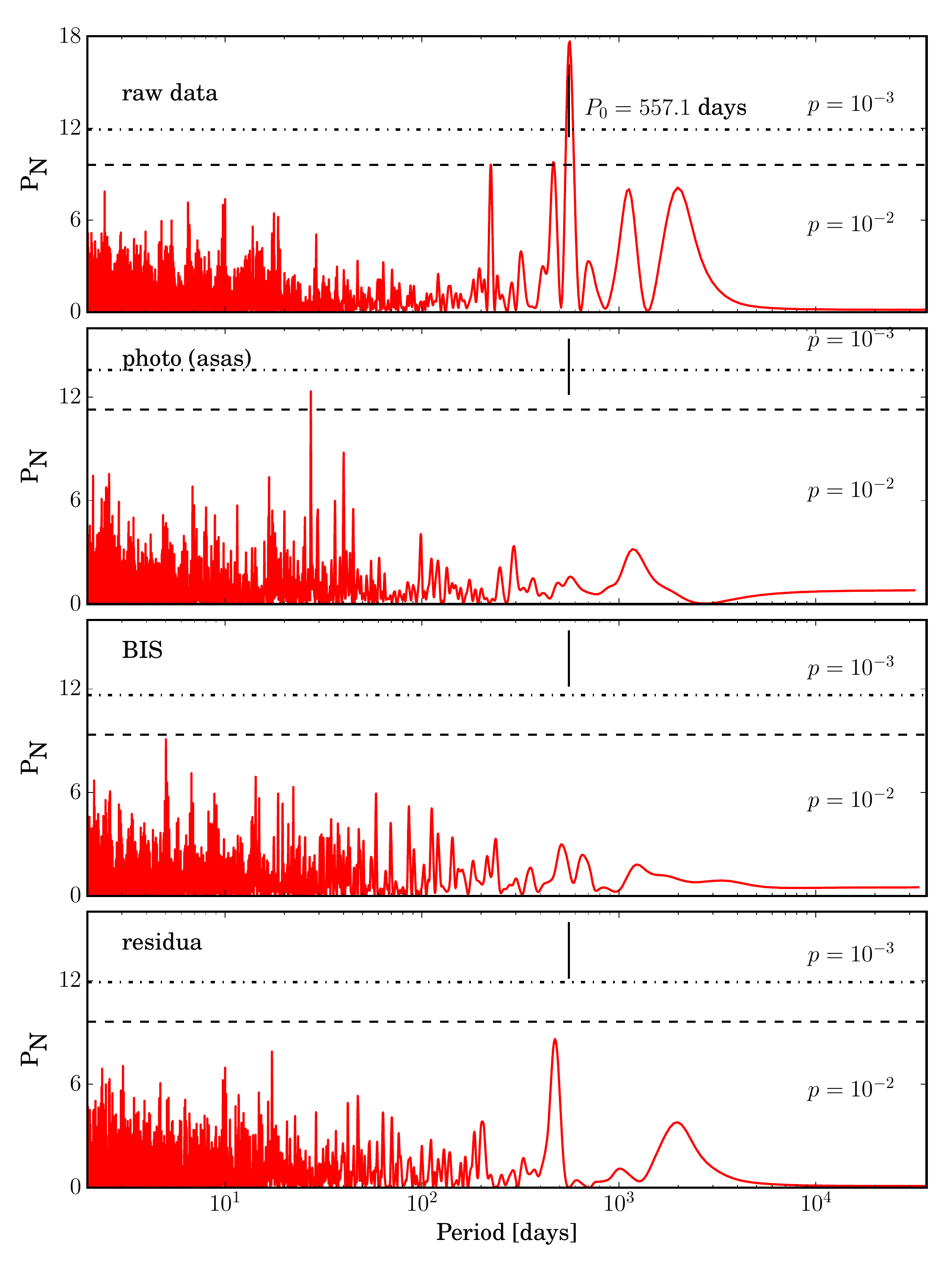}
   \caption{From top to bottom Lomb-Scarge periodograms for (a) the original HET HRS and HARPS-N  RV data of \starA, (b) ASAS photometry, (c) Bisector analysis,  and (d) RV residua
(HET and TNG) after the best Keplerian planet fit.}
   \label{LSP_1}
\end{figure}

For HD 103485 
two extensive sets of photometric observations are available from Hipparcos \citep{1997A&A...323L..49P, 2007A&A...474..653V} and ASAS \citep{1997AcA....47..467P}. 133 epochs of Hipparcos data were gathered over 1140 days between JD 2447878.4 and 2449019.0, long before our monitoring of this star. The average brightness is v$_{Hip}=8.422 \pm 0.013$ mag and they show no trace of variability. 
392 epochs of ASAS photometry were collected between JD 2451871.9 and 2455040.5 (3169 days), partly during our HET observations. These show average brightness of v$_{ASAS}=8.281\pm 0.01$ and trace of a 27d period, possibly due to the Moon. 
No significant photometric variability similar to that shown in RV is present in the available data as illustrated by the LS periodogram in Figure \ref{LSP_1}.

\begin{figure}
   \centering
   \includegraphics[width=0.5\textwidth]{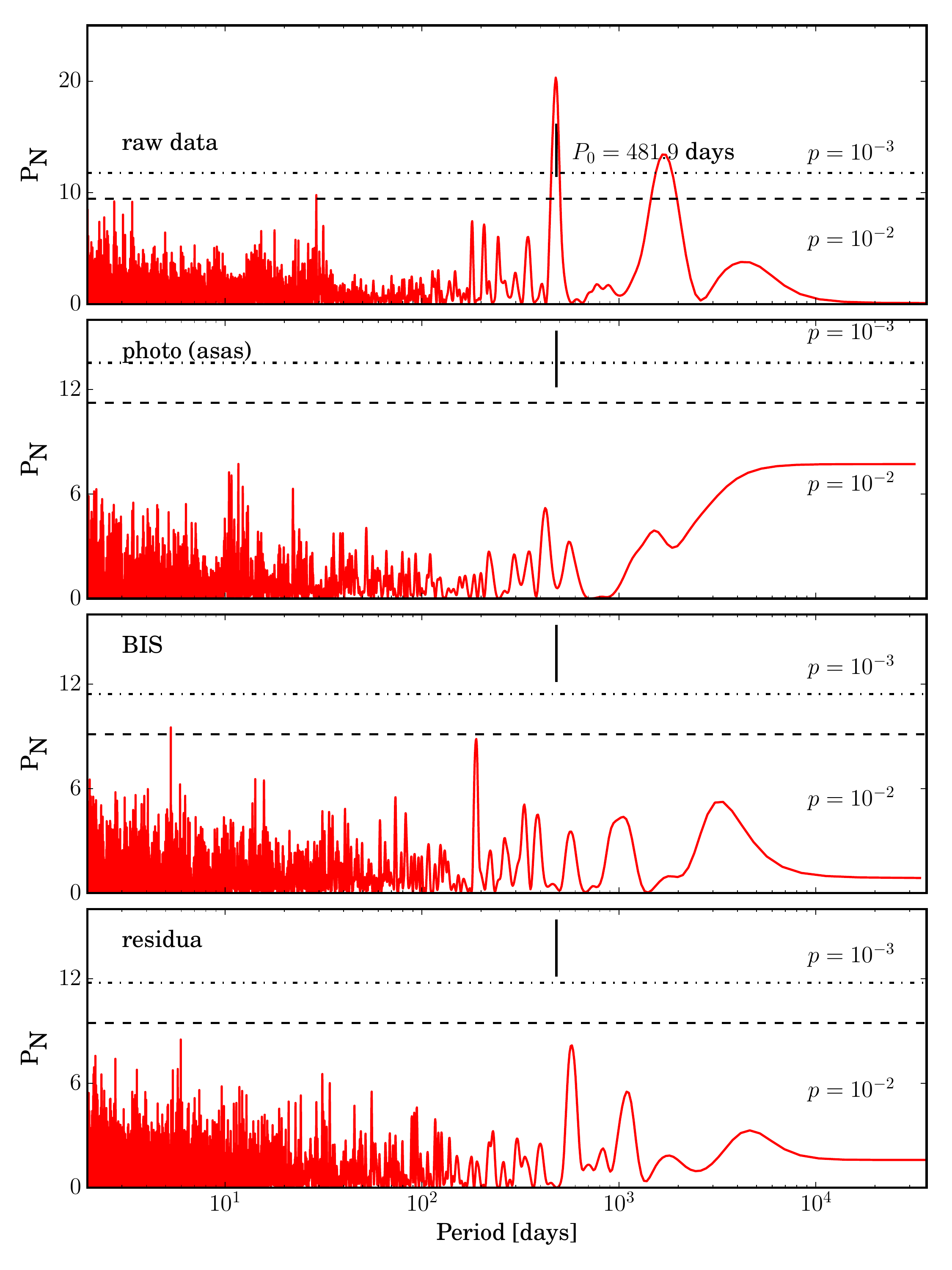}
   \caption{Same as Fig.\ref{LSP_1} for \starB}
   \label{LSP_2}
\end{figure}

BD+03 2562  was observed within ASAS  over 3169 days between JD 2451871.9 and 2455040.5, partly covering the timespan of our HET observations. The average brightness is v$_{ASAS}=9.497\pm0.015 $ mag and we find no trace of activity in these data (see Figure \ref{LSP_2}).

The spectral line bisectors and the calcium H$\&$K line shape show that we are dealing with a Keplerian motion that alters the position of the observed absorption lines in the spectra of both stars. Both weak and uncorrelated  variations of $\mathbf{H\alpha}$ and the lack of photometric variability in the case of \starB support that conclusion. In the case of \starA the $\mathbf{H\alpha}$ variations, weakly correlated with the observed RV in the TNG/HARPS-N data suggest that the observed RV variations may be due to a spot, but no such spot is visible in the photometric data, partly contemporaneous with our spectroscopic observations. 

We can therefore conclude that, although in the case of \starA the activity should be studied in more detail in the future, there exists no inexorable evidence that contradicts the interpretation of the observed RV variations as Doppler displacements due to the presence of a companion. 

\section{Keplerian analysis \label{results-g}}

\begin{figure}
   \centering
   \includegraphics[width=0.5\textwidth]{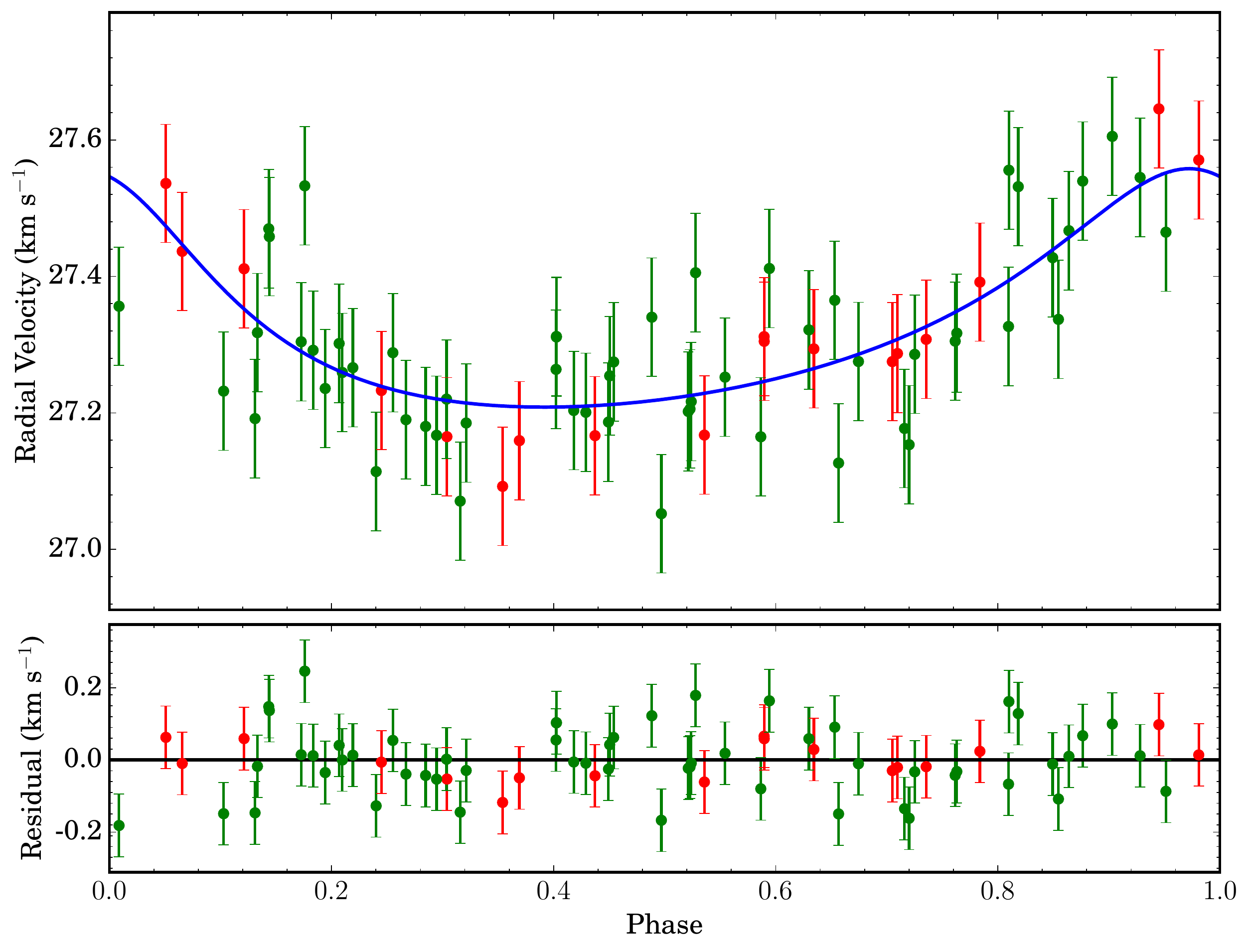}
   \caption{Keplerian best fit to combined HET HRS (green points) and TNG HARPS-N (red points) data for
      \starA. The estimated jitter due to p-mode oscillations has been added to the uncertainties.}
   \label{Fit_1}
\end{figure} 

\begin{figure}
   \centering
   \includegraphics[width=0.5\textwidth]{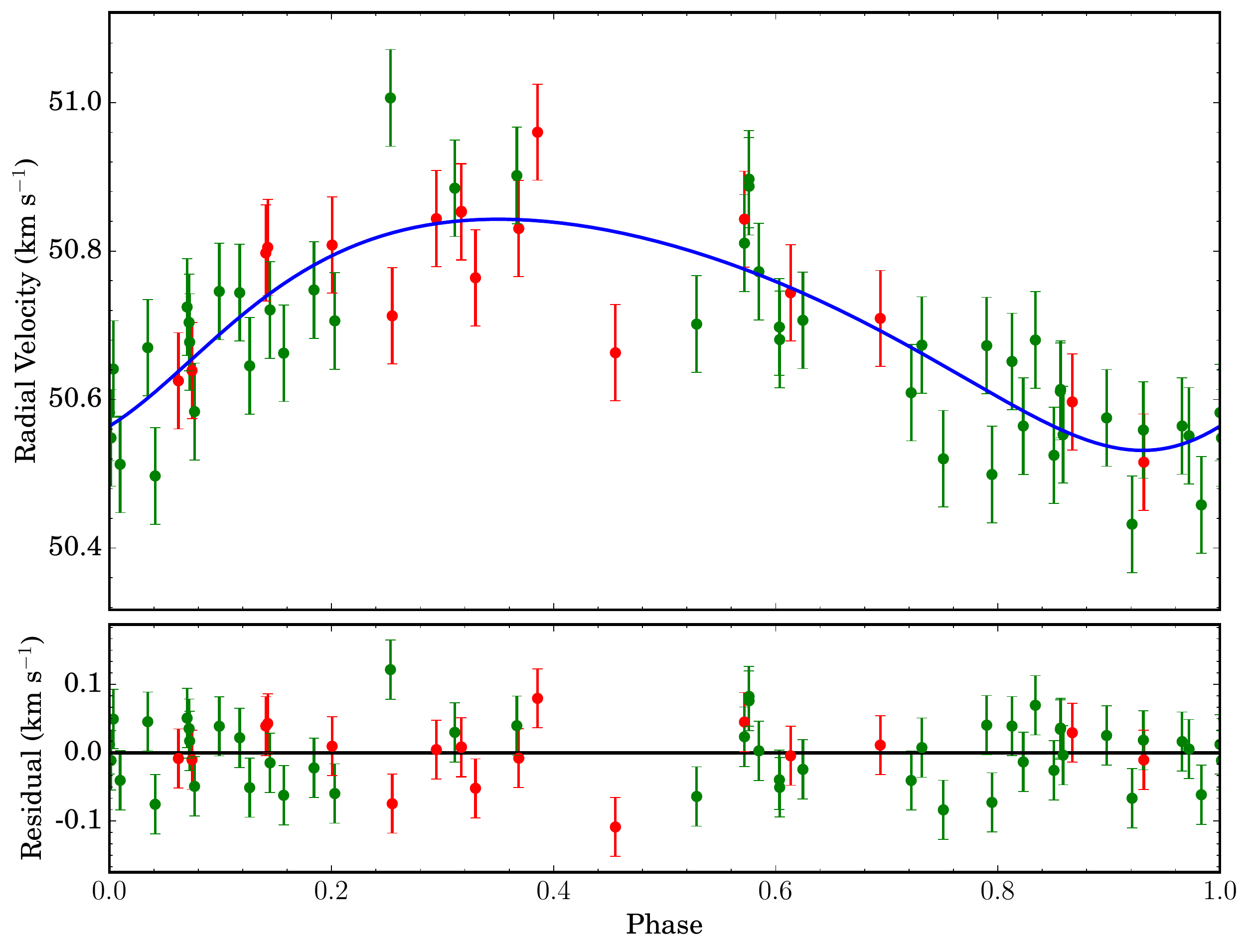}
   \caption{Keplerian best fit to combined HET HRS and TNG HARPS-N data for
      \starB. The jitter is added to the uncertainties.}
   \label{Fit_2}
\end{figure} 

The Keplerian orbital parameters have been derived using a hybrid approach (e.g.
\citealt{2003ApJ...594.1019G, 2006A&A...449.1219G, 2007ApJ...657..546G}), in
which  the PIKAIA-based, global genetic algorithm (GA;
\citealt{Charbonneau1995})  was combined with the
MPFit algorithm \citep{Markwardt2009}, to find the best-fit Keplerian orbit
delivered by RVLIN \citep{WrightHoward2009} modified  to allow the stellar
jitter to be fitted as a free parameter \citep{2007ASPC..371..189F,
2011ApJS..197...26J}.
The RV bootstrapping  method  \citep{1993ApJ...413..349M, 1997A&A...320..831K,
Marcy2005, Wright2007}  is employed to assess the uncertainties of the best-fit
orbital parameters (see TAPAS I for more details).  The results of the Keplerian
analysis for \starA are presented in Table \ref{KeplerianFit1} and in Figure \ref{Fit_1} and for \starB in 
Table \ref{KeplerianFit2} and in Figure \ref{Fit_2}. 

\begin{table}
\centering
\caption{Keplerian orbital parameters of \starA}
\renewcommand{\arraystretch}{1.3}
\begin{tabular}{ll}
\hline
Parameter & \starA  \\
\hline
\hline
$P$ (days)             & $557.1^{+5.0}_{-4.5}$  \\
$T_0$ (MJD)            & $53656^{+32}_{-35}$    \\
$K$ (\!\ms)            & $175^{+16}_{-15}$      \\
$e$                    & $0.34^{+0.16}_{-0.08}$ \\
$\omega$ (deg)         & $21^{+50}_{-50}$       \\
$m_2\sin i$ (\!\Mjup)  & $7 \pm 2$ \\
$a$ (\!\au)            & $1.4 \pm 0.1$ \\
$V_0$ (\!\ms)          & $27328.2^{+5.0}_{-5.2}$\\
offset (\!\ms)         & $27307^{+26}_{-27}$    \\
\sjit (\!\ms)  	     & $86.6$    \\
$\sqrt{\chi_\nu^2}$    & $1.05$    \\
RMS (\!\ms)            & $88.3$    \\
$N_{\textrm{obs}}$     & $75$      \\
\hline
\end{tabular}
\renewcommand{\arraystretch}{1}

\tablefoot{$V_0$ denotes absolute velocity of the barycenter of the system,
offset is a shift in radial velocity measurements between different telescopes,
\sjit~is stellar intrinsic jitter as defined in \cite{2011ApJS..197...26J},
RMS~is the root mean square of the residuals.}
\label{KeplerianFit1}
\end{table}

\begin{table}
\centering
\caption{Keplerian orbital parameters of \starB}
\renewcommand{\arraystretch}{1.3}
\begin{tabular}{ll}
\hline
Parameter & \starB  \\
\hline
\hline
$P$ (days)             & $481.9^{+2.7}_{-2.8}$  \\
$T_0$ (MJD)            & $53726^{+41}_{-32}$    \\
$K$ (\!\ms)            & $155.7^{+1.0}_{-3.3}$      \\
$e$                    & $0.20^{+0.12}_{-0.08}$ \\
$\omega$ (deg)         & $218^{+32}_{-26}$       \\
$m_2\sin i$ (\!\Mjup)  & $6.4 \pm 1.3$ \\
$a$ (\!\au)            & $1.3 \pm 0.1$ \\
$V_0$ (\!\ms)          & $50712^{+7}_{-5}$\\
offset (\!\ms)         & $50648^{+24}_{-24}$    \\
\sjit (\!\ms)  	     & $64.7$    \\
$\sqrt{\chi_\nu^2}$    & $1.11$    \\
RMS (\!\ms)            & $69.8$    \\
$N_{\textrm{obs}}$     & $64$      \\
\hline
\end{tabular}
\renewcommand{\arraystretch}{1}

\tablefoot{$V_0$ denotes absolute velocity of the barycenter of the system,
offset is a shift in radial velocity measurements between different telescopes,
\sjit~is stellar intrinsic jitter as defined in \cite{2011ApJS..197...26J},
RMS~is the root mean square of the residuals.}
\label{KeplerianFit2}
\end{table}

\section{Discussion and conclusions\label{conclusions}}
In this paper, the fifth of our TAPAS series, we present a planetary mass companion and a possible one to two very metal poor giant stars in the constellation of Virgo. We have interpreted the RV variations measured in both stars as Keplerian motion,  and while there is a lack of compelling evidence that the signal is  originated by stellar activity for \starB it is not so clear for \starA for which based on the available data both interpretations for the RV signal (planet and activity) are possible. In the meanwhile more data is gathered for \starA  it is hard to justify or disprove both possible interpretations so we keep as a working hypothesis for the following discussion that the RV signal is originated by a planet.  In this case, both giant stars have masses close to Solar (1.11 \Msun \starA\thinspace and 1.14 \Msun \starB), very low metallicities ($[Fe/H]= -0.50$ and $-0.71$ for \starA and \starB respectively) and  Jupiter planetary mass companions ($m_2\sin i\thinspace=\thinspace7$  and 6.4 \Mjup for \starA and \starB resp.) in close to terrestrial orbits (1.4~au \starA and 1.3~au \starB) with moderate eccentricities ($e=0.34$ and 0.2 for \starA  and \starB).

\begin{figure}
   \centering
   \includegraphics[width=0.5\textwidth]{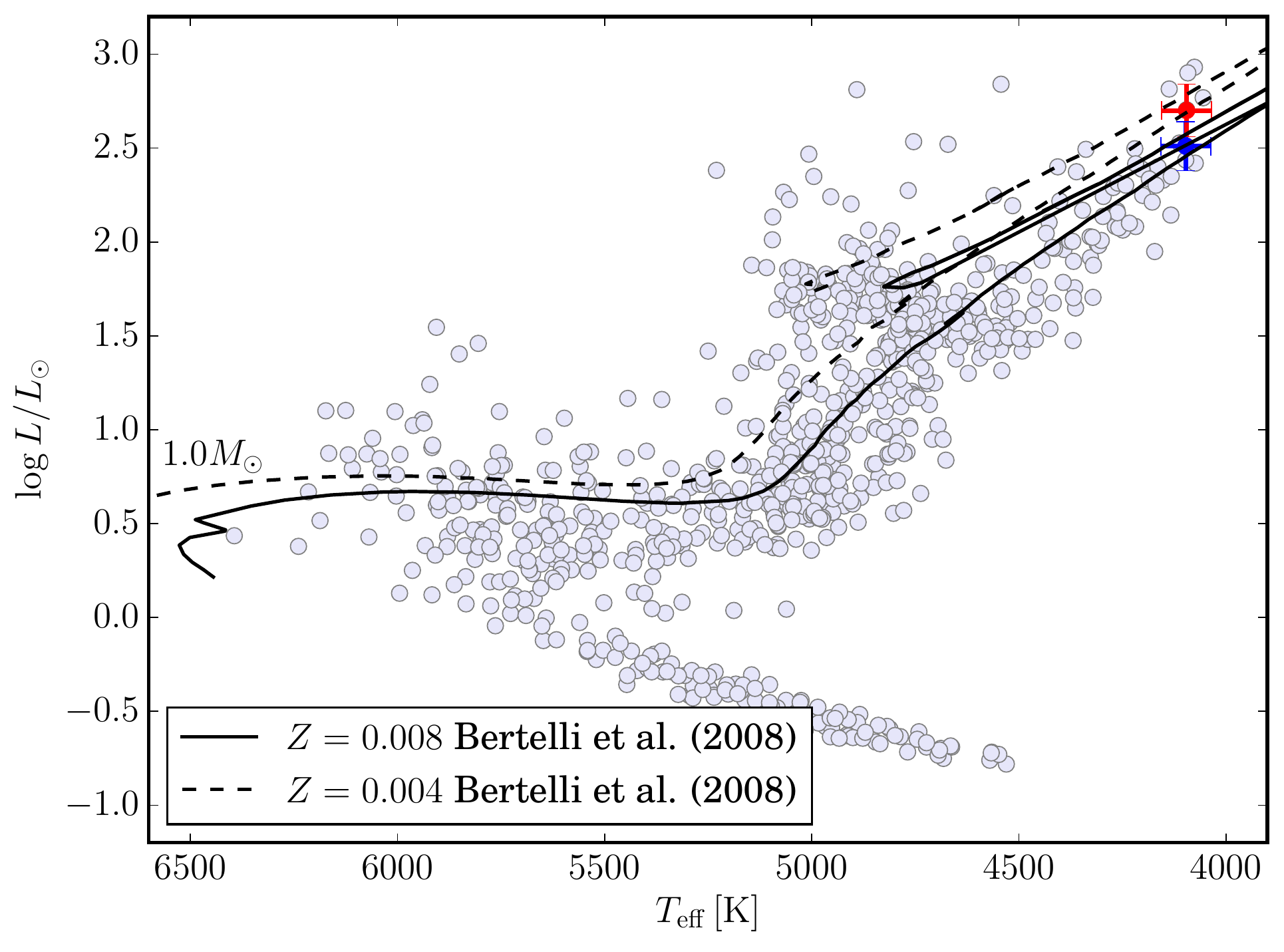}
   \caption{Hertzsprung-Russell  diagram  for  the  complete  PTPS  sample
with \starA (in blue) and \starB (in red) and the evolutionary tracks from 
\cite{Bertelli2008} for a star with 1 \Msun and metallicities $Z=0.008$ and $Z=0.004$ (see the legend in the bottom left corner of the plot).} 
   \label{Evo}
\end{figure}

In Fig.\thinspace\ref{Evo} we show the location of all the stars included in the PTPS sample on the Hertzsprung-Russel (HR) diagram where we have marked the location of \starA (blue) and \starB (red) (and their  corresponding uncertainties). The \cite{Bertelli2008} evolutionary tracks of a 1\Msun stars at two different very low metallicities are also shown for comparison. From the figure is clear that the two stars presented in this paper are among the most evolved stars of the whole PTPS sample. With ages of 6.17 and 5.25~Gyr for \starA and \starB (see Tables\ref{Parameters1} and \ref{Parameters2}), these stars are certainly above the mean age value of 3.37~Gyr obtained for Giant stars with planets in \cite{Maldonado2016}.

Again under the interpretation that the RV signal measured in both stars is due to Keplerian motion and based on the derived orbital parameters we compute the orbital solution under stellar evolution. None of the planets are expected to have experienced orbital decay caused by stellar tides at their current location (with $a/R_{\star}=10.99$ and 8.64 for \starA and \starB respectively) (see e.g. \thinspace \citealt{VillaverLivio2009,Villaver2014}). Given the tidal dissipation mode that operates in these giant stars, the planets should experience eccentricity decay together with orbital decay \citep{Villaver2014}, thus their moderate eccentricities and their $a/R_{\star}$ ratios are consistent with both planets being yet too far from the star to have experience tidal forces.  Both planets reported in this paper are located in similar regions in  $M_{\star}$ versus orbital distance or the $a-e$ plane as most of the other planets orbiting giant stars reported in the literature \citep{Villaver2014}. Neither \starA\thinspace b nor  \starB\thinspace b, are expected to survive engulfment when the star evolves up the tip of the RGB  \citep{VillaverLivio2009}. Assuming an average value of $\sin i$ both these companions stay within the planetary-mass range.

\begin{figure}
   \centering
   \includegraphics[width=0.5\textwidth]{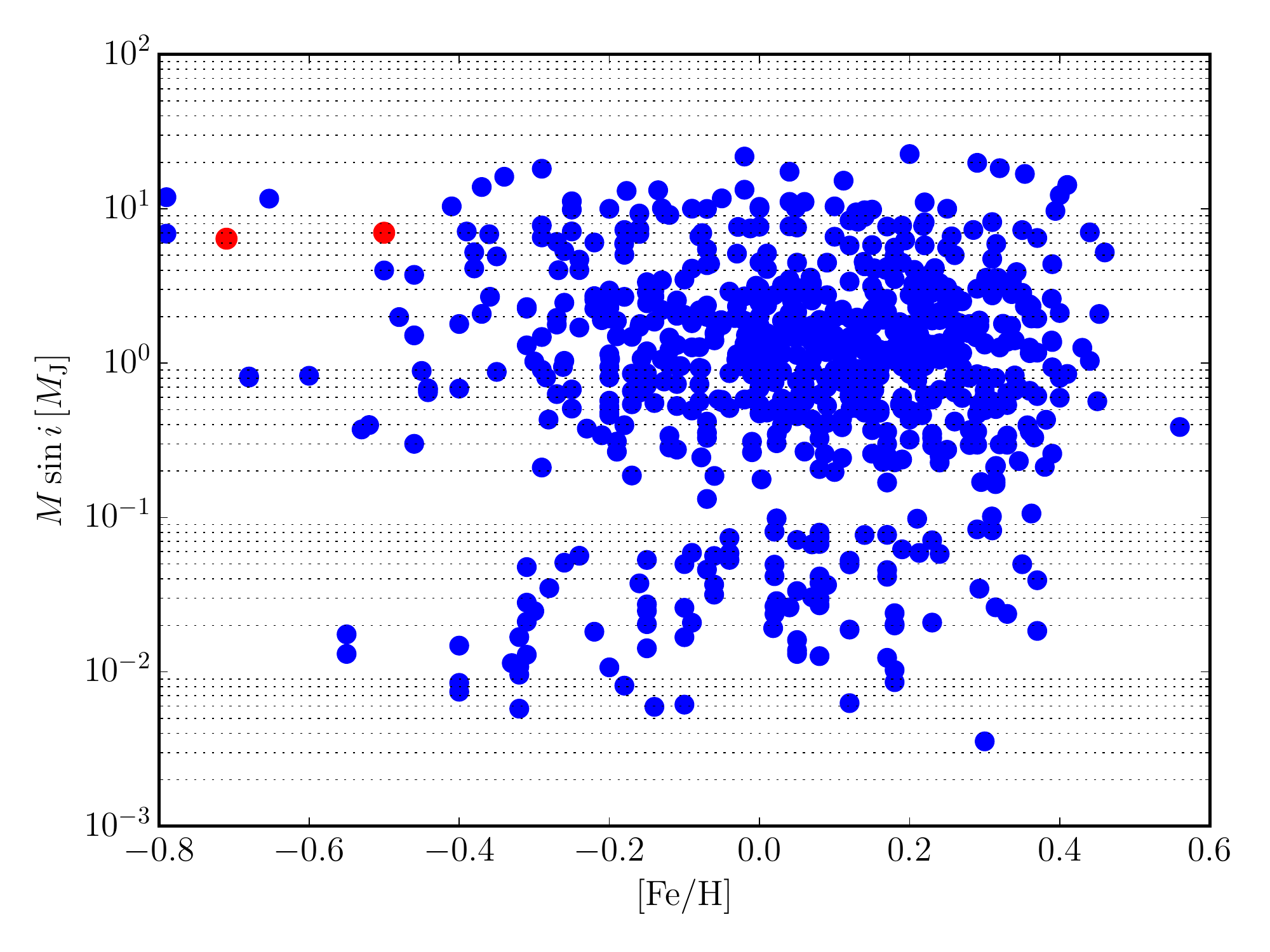}
   \caption{Planet minimum mass ($M \sin i$ in Jupiter mass) versus stellar metallicity ([Fe/H]) for all the confirmed planets (blue points) as taken from the Exoplanet encyclopedia (exoplanet.eu, exoplanets.org).  The two red points represent the location of the planets reported in this paper.  The horizontal lines are to guide the eye in the logarithmic scale in the vertical axis.}
\label{ZMp}   
\end{figure}

\begin{figure}
   \centering
   \includegraphics[width=0.5\textwidth]{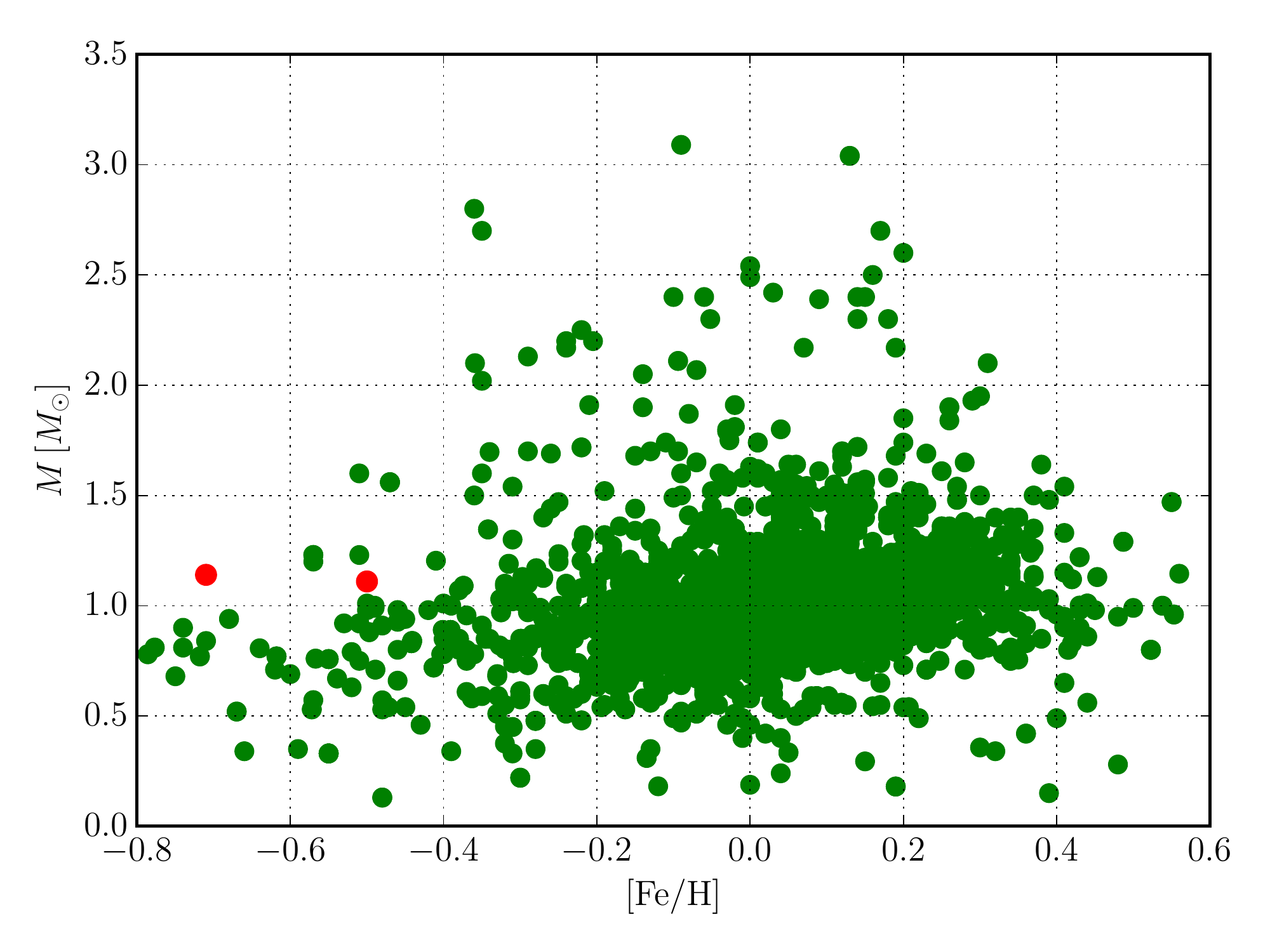}
   \caption{Stellar mass (in \Msun) versus stellar metallicity ([Fe/H]) for all the stars with confirmed planets (green points) according to the Exoplanet encyclopedia (exoplanet.eu, exoplanets.org).  The two red points represent the location of the stars reported in this paper. }
   \label{ZMs}
\end{figure}

Thus the orbital characteristics of the substellar object and the posible one we report in this paper do not appear to be different from the ones shown by the bulk population of planets found orbiting giant or subgiant stars. These two objects, however, clearly stand out in two important aspects: i) they are among the few very massive planets found around metal poor stars, and ii)  \starB in particular populates a region in the $M_{\star} -[\mathrm{Fe/H}]_{\star}$  plane where only another star has been found to host planets BD +20 2457  (see Figs.\thinspace \ref{ZMp} and \thinspace\ref{ZMs}).  

From Figs.\thinspace\ref{ZMp},\thinspace\ref{ZMs} it is clear that the two planet/star combinations reported in this paper are very special. First, they are two of the very few massive planets orbiting around stars close to the mass of the Sun to be found at very low metallicity. At lower metallicities than \starB, only two other planets have been reported in the literature orbiting the stars BD +20 2457  \citep{Niedzielski2009b} and  HD 11755  \citep{2015A&A...584A..79L}. The similarities among these systems are striking: giant, close to Solar mass, evolved stars with radii close to 30 \Rsun, with $\approx$ 7 \Mjup minimum mass planets and in  $\approx 1 au$ orbits.  Note also that the planet around \starA is the only massive planet in the region around stars with metallicity in the range $-0.6< [\mathrm{Fe/H}]_{\star} < -0.4$. \starA and \starB stand out even more in  Fig.\thinspace \ref{ZMs} where very few planets are known with  $[Fe/H]_{\star} < -0.48$ and $M_{\star}>1$\Msun.  

Our current understanding of massive planet formation offers two channels. First, core accretion $-$ and the growing of planets from the accretion of a gas envelope into a massive core $-$ has problems to easily explain systems formed at low metallicity.  
Models of planet formation by core accretion require a  protoplanetary disk with a high density of solids to form planetary cores which accrete gas before the primordial gas disk dissipates (see \thinspace\citealt{Ida2004}). The probability of a star hosting a planet that is detectable in radial velocity surveys increases 
$P_{pl} (Z) = 0.03 \times 10^{2\times Z}$, where $Z = [\mathrm{Fe/H}]$ is the stellar metallicity between $-0.5$ and 0.5~dex \citep{Gonzalez1997, FischerValenti2005}.  Thus although, core accretion does not exclude the formation at low metallicity, the probability of finding such planets is low $P_{pl}$ = 0.2\% for \starA and 0.1\% for \starB.  

The alternative mechanism,  in-situ fragmentation via gravitational instability  (GI)  (see e.g., \citealt{Cameron1978,Boss1997}) posses such strong requirements in the characteristics of a protoplanetary disk at $1\thinspace \mathrm{au}$  that has been shown cannot exit on dynamical grounds (see e.g. \citealt{Rafikov2005, Stamatellos2008}). The planet could form by GI at larger distances \citep{Rafikov2005,Matzner2005} and experience subsequent migration to $\approx 400 d$ orbital periods in relatively short timescales. In fact, the fragments formed by instability at $100\thinspace\mathrm{au}$  are expected to have minimum masses  of $10 \Mjup$ \citep{Rafikov2005} which is approximately the typical mass of the planets found around the low metallicity stars reported in this paper. Note that  although it has been shown that GI cannot be the main channel for planet formation as it cannot reproduce the overall characteristics of the bulk of the planet detections, nothing prevents it to be the preferred mechanism under certain circumstances.  In fact, it has been reported that protoplanetary disks with low metallicities generally cool faster and show  stronger overall GI activity \citep{Mejia2005,Cai2006} although the lowest metallicity consider in these models is a quarter Solar (still much larger than the ones shown by the stars presented in this paper). So the question still remains on how gas cooling in the disk operates at the low metallicity of these stars given the disk needs to be atypically cold for GI and whether these planets represent indeed the low-mass tail of the distribution of disk-born companions \citep{Kratter2010}. 

Planet host giants, in fact, have been reported to show peculiar characteristics regarding the planet-metallicity relation. In particular, \cite{Maldonado2013} show that, whilst the metallicity distribution
of planet-hosting giant stars with stellar masses $M > 1.5\Msun$ follows the general trend that has been established for main sequence stars hosting planets, giant planet hosts in the mass domain $M \leqslant 1.5\Msun$ do not show metal enrichment. Similar results were found by \cite{Mortier2013}.  Note that \cite{Reffert2015} challenged these results based on a discussion of planet contamination but it has been shown by\cite{Maldonado2016} based on their planet list that the result is sustained using the \cite{Reffert2015} list of candidates.

The two objects presented in this paper add two more points to a already puzzling relation between giant planets and giant stars that might help understand planet formation mechanisms for low metallicity stars.   

\begin{acknowledgements}
We thank the HET and IAC resident astronomers and telescope operators  for
their support.
EV acknowledges support from the Spanish Ministerio de Econom\'ia y
Competitividad under grant AYA2014-55840P.

MA acknowledges the Mobility+III fellowship from the Polish Ministry of Science
and Higher Education. 

AN, BD-S and MiA  were supported by the Polish National Science Centre grant
no. UMO-2012/07/B/ST9/04415 and UMO-2015/19/B/ST9/02937.

KK was funded in part by the Gordon and Betty Moore Foundation's
Data-Driven Discovery Initiative through Grant GBMF4561.

This research was supported in part by PL-Grid Infrastructure.

The HET is a joint project of the University of Texas at Austin, the
Pennsylvania State University, Stanford University, Ludwig-
Maximilians-Universit\"at M\"unchen, and Georg-August-Universit\"at
G\"ottingen. The HET is named in honor of its principal benefactors, William P.
Hobby and Robert E. Eberly.  The Center for Exoplanets and Habitable Worlds is
supported by the Pennsylvania State University, the Eberly College of Science,
and the Pennsylvania Space Grant Consortium.

This work made use of NumPy~\citep{numpy}, Matplotlib~\citep{mpl},
Pandas~\citep{pandas} and \texttt{yt}~\citep{yt} and of the Exoplanet Orbit Database and the ExoplanetData Explorer at exoplanets.org and exoplanet.eu. 

\end{acknowledgements}
\bibliographystyle{aa} 
\bibliography{villaver} 

\newpage

%
\begin{table}
\centering
\caption{HET and HRS RV and BIS measurements  (\!\ms) of \starA}
\begin{tabular}{lrrrr}
\hline
MJD& RV  & \srv & BIS  & \sbs \\
\hline\hline
53023.396748 & 159.57 & 6.67 & 52.97 & 20.25 \\
53349.497471 & -120.73 & 6.97 & 18.92 & 23.61 \\
53389.521123 & -105.25 & 9.38 & 0.94 & 9.39 \\
53390.398189 & -101.55 & 6.63 & 5.59 & 20.29 \\
53713.504688 & -75.47 & 5.03 & 61.34 & 16.20 \\
53730.469410 & 10.27 & 5.01 & 17.97 & 15.08 \\
53736.450347 & 150.99 & 6.23 & 4.91 & 23.15 \\
53752.421736 & -3.16 & 5.77 & 38.96 & 19.03 \\
53758.387778 & -15.56 & 4.78 & 85.74 & 13.41 \\
53764.472211 & -71.56 & 5.11 & 11.97 & 15.25 \\
53771.468009 & -5.62 & 5.90 & 30.50 & 21.04 \\
53778.334097 & -41.08 & 5.14 & 57.19 & 13.22 \\
53798.402685 & -19.16 & 5.26 & 35.25 & 13.23 \\
53820.327072 & -140.15 & 5.65 & 39.07 & 14.60 \\
53825.333646 & -87.17 & 6.98 & 24.80 & 25.16 \\
53832.184578 & -236.60 & 5.06 & 30.82 & 11.02 \\
53835.190376 & -122.14 & 4.87 & 2.87 & 11.36 \\
53889.142350 & -103.83 & 5.78 & 47.98 & 13.38 \\
54080.494797 & -2.24 & 5.26 & -14.81 & 21.43 \\
54107.425972 & 248.28 & 5.39 & -18.00 & 15.35 \\
54129.379907 & 120.04 & 5.37 & 85.30 & 13.44 \\
54144.454317 & 232.28 & 5.22 & 72.06 & 17.44 \\
54159.302743 & 297.81 & 4.65 & 40.43 & 14.48 \\
54173.254462 & 237.63 & 4.84 & 63.52 & 16.70 \\
54186.212963 & 157.43 & 4.50 & 78.54 & 11.76 \\
54218.127784 & 48.89 & 5.49 & 31.04 & 15.28 \\
54437.514450 & 3.88 & 6.08 & -4.03 & 18.40 \\
54437.520914 & 4.68 & 5.09 & 12.48 & 12.34 \\
54485.392633 & 32.88 & 5.49 & 46.55 & 18.89 \\
54507.337703 & 98.05 & 6.34 & -7.39 & 19.88 \\
54544.358096 & 104.23 & 5.67 & 2.48 & 15.79 \\
54564.184074 & 14.31 & 5.95 & 16.64 & 14.78 \\
54577.150104 & 57.68 & 5.80 & 40.09 & 16.05 \\
54612.164201 & -130.14 & 6.93 & 25.87 & 20.25 \\
54843.425116 & -115.73 & 5.87 & -9.33 & 10.13 \\
54868.351563 & 225.36 & 5.32 & 28.37 & 16.97 \\
55171.525139 & -153.91 & 5.35 & 44.17 & 17.87 \\
55195.453252 & 9.48 & 5.09 & 2.68 & 16.53 \\
55221.392708 & 19.29 & 5.05 & 7.21 & 16.05 \\
55246.422118 & 29.68 & 5.25 & 72.43 & 14.38 \\
55551.487361 & -43.54 & 6.12 & 9.95 & 14.16 \\
55566.435694 & -106.47 & 4.97 & -2.00 & 16.87 \\
55578.398380 & -53.02 & 5.03 & 77.56 & 14.94 \\
55580.392459 & -32.69 & 6.60 & 63.22 & 21.96 \\
55604.322911 & -255.04 & 4.88 & 61.88 & 15.32 \\
55619.301128 & -90.72 & 5.34 & 47.17 & 16.51 \\
55636.263281 & -54.98 & 5.54 & 11.47 & 17.44 \\
55654.195978 & -142.37 & 5.81 & 44.86 & 11.18 \\
55693.220660 & -180.79 & 6.35 & 47.69 & 15.09 \\
55703.188322 & -32.04 & 6.42 & 42.51 & 12.06 \\
55964.447500 & 162.42 & 4.27 & 19.69 & 10.58 \\
56001.356383 & -48.11 & 4.70 & 51.40 & 12.81 \\
56018.315793 & -193.32 & 5.35 & 4.04 & 12.91 \\
56033.268374 & -117.32 & 5.65 & 49.77 & 15.72 \\
56043.131134 & -127.15 & 6.03 & 42.22 & 15.92 \\
56288.476146 & -21.54 & 4.70 & 52.59 & 16.50 \\
56340.424410 & 224.16 & 4.73 & 68.38 & 14.27 \\
\hline
\end{tabular}

\label{HETdata1}
\end{table}
%
%
\begin{table}
\centering
\caption{TNG and HARPS-N RV and BIS measurements (\!\ms) of \starA}
\begin{tabular}{lrrr}
\hline
MJD& RV  & \srv & BIS  \\
\hline\hline
56277.247179 & 27275.10 & 1.97 & 157.38 \\
56294.234787 & 27307.70 & 1.20 & 198.41 \\
56321.174370 & 27391.60 & 1.10 & 186.65 \\
56410.966775 & 27645.50 & 1.35 & 197.62 \\
56430.962559 & 27570.60 & 1.02 & 210.71 \\
56469.909044 & 27536.20 & 1.43 & 209.60 \\
56647.244692 & 27159.30 & 3.09 & 185.54 \\
56685.175862 & 27166.60 & 1.70 & 192.14 \\
56740.046309 & 27167.50 & 2.14 & 226.24 \\
56770.016852 & 27311.80 & 1.57 & 209.44 \\
56770.065133 & 27304.80 & 1.24 & 206.05 \\
56794.966883 & 27293.80 & 2.02 & 221.75 \\
56836.907914 & 27287.00 & 2.06 & 229.18 \\
57035.192541 & 27436.70 & 2.22 & 234.18 \\
57066.212695 & 27411.20 & 3.26 & 240.97 \\
57135.092065 & 27232.80 & 1.36 & 220.46 \\
57167.958789 & 27165.10 & 0.65 & 242.70 \\
57195.906134 & 27092.30 & 1.31 & 213.33 \\
\hline
\end{tabular}

\label{HARPSdata1}
\end{table}

\begin{table}
\centering
\caption{HET and HRS RV and BIS measurements  (\!\ms) of \starB}
\begin{tabular}{lrrrr}
\hline
MJD& RV  & \srv & BIS  & \sbs \\
\hline\hline
53037.367523 & 162.58 & 9.31 & 59.79 & 25.13 \\
53039.337060 & 248.83 & 10.61 & 81.11 & 32.98 \\
53039.350613 & 239.24 & 11.03 & 57.66 & 29.48 \\
53341.515486 & 58.00 & 7.32 & 38.91 & 12.23 \\
53759.367361 & 76.58 & 6.54 & 40.39 & 18.48 \\
53773.326875 & 97.62 & 6.06 & 39.61 & 22.59 \\
53801.268322 & 14.42 & 6.09 & 47.88 & 19.55 \\
54127.490388 & 32.19 & 7.26 & 70.01 & 24.44 \\
54138.338883 & -37.01 & 8.54 & 61.44 & 20.41 \\
54138.450851 & -34.41 & 8.24 & 86.82 & 22.30 \\
54158.399971 & -72.73 & 7.45 & 38.77 & 26.63 \\
54174.364925 & -89.03 & 5.57 & 41.44 & 17.05 \\
54191.197078 & -83.65 & 6.55 & 55.70 & 16.84 \\
54194.174068 & -96.84 & 6.35 & 39.17 & 20.00 \\
54208.265365 & -99.68 & 7.78 & 55.13 & 22.80 \\
54209.261464 & -6.83 & 6.99 & 46.65 & 26.47 \\
54212.253084 & -135.25 & 6.87 & 45.89 & 24.86 \\
54224.223663 & 21.94 & 5.59 & 64.43 & 18.52 \\
54242.173779 & 55.84 & 6.93 & -40.03 & 21.22 \\
54264.123675 & 95.88 & 7.55 & 31.74 & 20.84 \\
54462.444896 & 53.61 & 7.03 & 30.05 & 23.78 \\
54498.347031 & 49.62 & 7.49 & -7.74 & 25.73 \\
54498.467564 & 32.75 & 7.85 & 36.94 & 26.35 \\
54560.185220 & 25.32 & 7.53 & 18.94 & 24.50 \\
54604.187876 & -83.64 & 7.62 & -12.55 & 21.49 \\
54811.497899 & 358.20 & 7.39 & 14.09 & 26.62 \\
54839.418223 & 236.66 & 6.98 & -0.96 & 13.58 \\
54866.333733 & 253.73 & 7.64 & 85.91 & 28.11 \\
55171.502292 & -65.61 & 6.18 & -3.30 & 22.57 \\
55208.423212 & -64.20 & 8.23 & -7.78 & 38.24 \\
55232.342153 & -2.61 & 5.44 & 45.44 & 17.90 \\
55260.257106 & 99.62 & 7.65 & 45.93 & 22.58 \\
55554.472876 & -148.97 & 7.14 & -43.34 & 29.10 \\
55581.385388 & -123.10 & 6.05 & 3.91 & 21.40 \\
55585.369132 & -95.27 & 5.93 & -46.36 & 20.95 \\
55615.289612 & -215.92 & 7.42 & 45.46 & 17.89 \\
55645.354259 & -189.93 & 6.51 & 28.50 & 21.04 \\
55673.272147 & -150.91 & 6.20 & 50.51 & 17.87 \\
55688.233883 & 29.30 & 6.69 & 41.05 & 20.16 \\
55723.130874 & 72.67 & 7.03 & 40.59 & 19.77 \\
55954.374120 & 58.71 & 6.46 & 31.59 & 22.22 \\
56001.366777 & -38.82 & 5.36 & 23.81 & 14.44 \\
56015.332512 & -127.79 & 7.08 & 41.77 & 20.97 \\
56034.152297 & 24.60 & 6.97 & 57.35 & 16.39 \\
56045.120087 & 3.30 & 6.22 & 66.61 & 16.64 \\
56417.215810 & 124.19 & 6.47 & 90.08 & 21.60 \\
\hline
\end{tabular}

\label{HETdata2}
\end{table}

\begin{table}
\centering
\caption{TNG and HARPS-N RV and BIS measurements (\!\ms) of \starB}
\begin{tabular}{lrrr}
\hline
MJD& RV  & \srv & BIS  \\
\hline\hline
56277.252412 & 50843.70 & 1.93 & 140.84 \\
56294.229018 & 50764.00 & 1.13 & 176.94 \\
56321.169034 & 50960.10 & 1.03 & 190.92 \\
56410.961652 & 50842.90 & 1.27 & 148.93 \\
56430.957199 & 50743.80 & 1.39 & 155.83 \\
56469.901805 & 50709.30 & 1.51 & 162.17 \\
56647.240508 & 50625.20 & 2.99 & 133.03 \\
56685.171295 & 50797.30 & 2.06 & 136.22 \\
56740.042885 & 50712.70 & 2.97 & 163.36 \\
56770.012446 & 50852.50 & 1.91 & 157.46 \\
56770.060731 & 50853.50 & 1.51 & 154.18 \\
56794.960756 & 50830.40 & 2.34 & 169.21 \\
56836.901346 & 50663.00 & 3.12 & 173.94 \\
57035.183363 & 50596.80 & 2.35 & 183.76 \\
57066.205983 & 50515.60 & 6.44 & 132.81 \\
57135.084365 & 50639.10 & 1.40 & 182.09 \\
57167.950415 & 50805.00 & 1.37 & 161.54 \\
57195.899321 & 50808.10 & 1.49 & 151.10 \\
\hline
\end{tabular}

\label{HARPSdata2}
\end{table}

\end{document}